\documentclass[runningheads]{llncs}
\usepackage{graphicx}

\usepackage{tikz}
\usepackage{comment}
\usepackage{amsmath,amssymb} 
\usepackage{color}
\usepackage{subfigure}
\usepackage{multirow}
\usepackage{float}
\usepackage{bbding}
\usepackage[accsupp]{axessibility}  

\begin{document}
\pagestyle{headings}
\mainmatter
\def\ECCVSubNumber{3030}  

\title{SeCo: Separating Unknown Musical Visual Sounds
with Consistency Guidance} 

\titlerunning{SeCo}
%
\author{Xinchi Zhou\inst{1}\thanks{Equal contribution.} \and
Dongzhan Zhou \inst{1}$^\star$ \and
Wanli Ouyang \inst{1} \and 
Hang Zhou \inst{2} \and
Ziwei Liu \inst{3} \and
Di Hu \inst{4}}
\authorrunning{Zhou et al.}
%
\institute{The University of Sydney\and
Baidu Inc. \and
S-lab, Nanyang Technological University \and
Gaoling School of Artificial Intelligence, Renmin University of China\\}
\maketitle

\begin{abstract}
Recent years have witnessed the success of deep learning on the visual sound separation task. However, existing works follow similar settings where the training and testing datasets share the same musical instrument categories, which to some extent limits the versatility of this task. In this work, we focus on a more general and challenging scenario, namely the separation of unknown musical instruments, where the categories in training and testing phases have no overlap with each other. To tackle this new setting, we propose the ``Separation-with-Consistency"~(SeCo) framework, which can accomplish the separation on unknown categories by exploiting the consistency constraints. 
Furthermore, to capture richer characteristics of the novel melodies, we devise an online matching strategy, which can bring stable enhancements with no cost of extra parameters. Experiments demonstrate that our SeCo framework exhibits strong adaptation ability on the novel musical categories and outperforms the baseline methods by a significant margin.
\end{abstract}

\section{Introduction}
The objective of visual sound separation is to separate the mixed audio signals into individual components with the guidance of visual cues. Deep neural networks can extract rich semantic information from both visual and auditory modalities, which significantly promote the development of the visual sound separation task. Most deep-learning based approaches, such as~\cite{Zhao_2018_ECCV,gao2019co,gan2020music}, adopt the setting where the training and testing sets share the same musical instrument categories. Despite the success, there still exist some limitations on such training mode, which confine the separation targets to the musical instruments that have appeared in the training set. The more general setting of visual sound separation on unknown musical instruments remains an unexplored problem.

In this work, we undertake the task of visually guided music separation on unknown classes, that is, the categories of training and testing sets have~\textit{\textbf{no overlap}}. 
Under the real scenario, it is challenging to directly identify and separate the sound of the unfamiliar musical instruments from the mixed audio signal, even for humans. Thus, it may be difficult to directly apply the existing frameworks to this new setting and reasonable priors must be added to enhance the adaptability of the deep models to the unknown musical sounds. 

To handle this challenging new setting, we propose a novel `Separation with Consistency' (SeCo) framework, which exploits the consistency constraints to realize the visual sound separation on the novel musical instruments. The system receives two types of consistency supervisions, namely the \textbf{inter-modal} consistency and the \textbf{intra-modal} consistency. Firstly, the audio-visual associations in videos are natural and will not change with different categories. Therefore, it is critical to strengthen the audio-visual (AV) correlations during training, instead of simply capturing the isolated features for the auditory and visual modalities. In this way, the visual cues can provide better separation guidance even for the categories that have never been seen before. 
Specifically, we require that the separated audio signals should be aligned with the visual components in the original videos, which is denoted as the inter-modal consistency. Secondly, even though directly identifying unfamiliar sounds is not easy for humans, things will become quite different if some auditory examples are provided. In particular, human brains can achieve the goal of separating the novel target sound by perceiving the similarities and differences between the mixed sound and the given template sound. Thus, it is a natural idea to incorporate such a template learning mechanism in deep models for the assistance of visual sound separation on unknown classes. Specifically, since the sounds from the same type of musical instruments normally enjoy similar timbres and tones, the features extracted from them should be close in the embedding space. In this way, the intra-modal consistency expands the supervision scale from the sample level to the wider category level, which can effectively help the deep models adapt well to the unknown classes.

For an unfamiliar melody, humans may listen to the melody over and over again to better capture the characteristics of the musical tone. Similarly, this behavior can also be applied to our task. We develop an online matching strategy to iteratively refine the predicted mask for each sample independently so that the potential of the devised consistency guidance can be further exploited. The online matching strategy can bring stable improvements without introducing any extra parameters.

Our \textbf{contributions} can be summarized as followed. (1) We explore the task of visual music separation under the scene of unknown musical instruments, which expands the scope of visual sound separation and makes the task more versatile. (2) We propose a novel framework, SeCo, to adapt to this challenging situation. The results show that our approach outperforms the baselines by a noticeable margin. We also conduct in-depth ablation studies to analyze the effects of the key parameters. (3) We design an online matching strategy, which brings consistent improvements with no extra parameter costs.

\section{Related Work}

\noindent \textbf{Audio-Visual Learning.} With the development of deep learning, audio-visual learning has also received widespread attention in recent years and breakthroughs have been made in various sub-fields~\cite{zhu2021deep}. Audio-visual representation learning aims at finding the correlations between the audio and visual modality in a self-supervised manner and thus provides good audio/visual representations~\cite{arandjelovic2017look,arandjelovic2018objects,owens2018learning,korbar2018cooperative,owens2018audio,hu2019deep}. In addition, many works utilize audio information to improve the video analysis tasks~\cite{kazakos2019epic,gao2020listen}. The objective of the audio-visual localization task is to localize the sound source in the visual context~\cite{senocak2018learning,Zhao_2018_ECCV,arandjelovic2018objects,hu2020discriminative}. Another important branch of the audio-visual learning field lies in the cross-modality generation, which consists of visual-to-audio~\cite{zhou2018visual,morgado2018self} and audio-to-visual~\cite{chung2017you,chen2018lip,ginosar2019learning,zhou2019talking} tasks. 
Most previous works in audio-visual learning require that the training and validation data come from the same domain or similar scenario, while our work investigates a more challenging setting where the training and testing set has no category overlaps.

\noindent \textbf{Visual Sound Separation.} By leveraging visual modality to the sound separation task, models can utilize the richer context information, which outperforms the single modality approaches. Visual sound separation is explored on various identities, such as speakers~\cite{gabbay2018seeing,ephrat2018looking,lu2018listen}, objects~\cite{gao2018learning} and musical sounds~\cite{Zhao_2018_ECCV,xu2019recursive,gao2019co,xu2019recursive,zhao2019sound,gan2020music}. Our work concentrates on the branch of visual music separation. 

Zhao et al.~\cite{Zhao_2018_ECCV} propose the PixelPlayer framework with the `mix-and-separation' paradigm, which learns to separate mixed audios into components and locate the sound production regions on images in a self-supervised manner. 
Considering the limitations of static images, many works attempt to adopt visual cues from other modalities and further benefit the separation task, such as motion~\cite{zhao2019sound}, skeleton~\cite{gan2020music} and scene graph~\cite{chatterjee2021visual}. Gao et al.~\cite{gao2021visualvoice} incorporate audio-visual consistency in the speech separation framework but the training and validation sets act on the same category, i.e., speakers. Unlike any of the above, we focus on the visual sound separation task of unknown musical instruments and also propose an effective framework to handle this new scene. 

\noindent \textbf{Transfer Learning on Novel Category.} Humans naturally have a strong ability to establish the perception of new objects, even with very limited samples. However, in most cases, machines can obtain such perception ability only if it has been fed enough examples. Thus, few-shot~\cite{fei2006one} and zero-shot~\cite{lampert2009learning} learning are proposed to investigate the transferability of machines when very few or even no samples are provided on new objects. Such learning paradigms can effectively reduce the burdens of data acquisition and storage. Few-shot learning approaches utilize the information of the limited samples from the new category, while the models have no exposure to any instances of the target class under the zero-shot setting. The mainstream solutions for the few-shot setting include metric learning~\cite{koch2015siamese,snell2017prototypical,sung2018learning} and meta-learning~\cite{finn2017meta,nichol2018first}. 
Zero-shot learning approaches usually transfer knowledge from familiar classes, such as semantic embedding~\cite{lampert2013attribute,kodirov2017semantic}, or exploit external information such as knowledge graphs~\cite{wang2018zero}. In addition to the classification scene, many works extend the few/zero-shot setting to other sub-fields such as object detection~\cite{kang2019few} and semantic segmentation~\cite{wang2019panet}. Despite the success in vision fields, it is still challenging to deploy novel category transfer learning on multi-modality models. By leveraging the inter-modal and intra-modal consistency guidance, our SeCo framework exhibits impressive transferring performance on unknown musical separation and serves as a strong baseline for this novel and challenging task.

\section{Methodology}
\begin{figure}[t]
    \centering
    \includegraphics[width=0.92\textwidth]{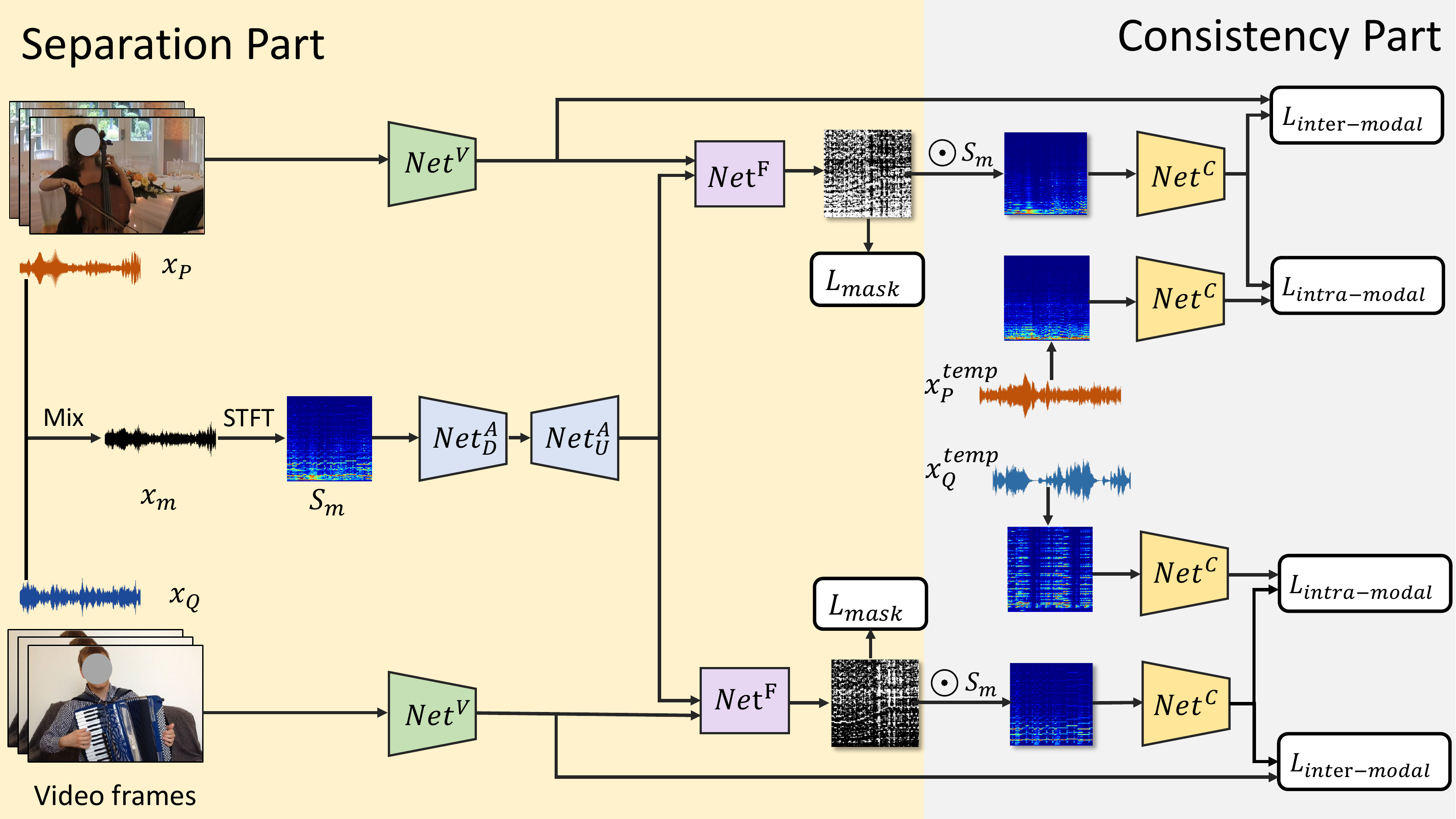}
    \caption{The whole pipeline of our ``Separation-with-Consistency" framework is composed of the separation part and the consistency part. In the separation stage, the visual features and audio features are extracted by the vision network $Net^V$ and audio network $Net^A$, respectively, and get fused in the fusion network $Net^{F}$ to predict the separation masks. In the consistency stage, the separated spectrograms and the template spectrograms pass through the consistency network $Net^{C}$ to generate the high-level features for the computation of the consistency constraints. The system is trained by minimizing the combination of the separation loss ($L_{mask}$) and the consistency loss ($L_{intra-modal}$ \& $L_{inter-modal}$).}
    \label{fig:pipeline}
\end{figure}

We propose the `Separation-with-Consistency' paradigm (SeCo) to achieve the transfer learning of visual sound separation on novel musical instruments. Specifically, the consistency guidance is composed of the inter-modal and intra-modal parts, which require the separated sounds to align with the corresponding visual cues and sounds of the same category. The pipeline of our SeCo framework is illustrated in Fig.~\ref{fig:pipeline}. We will first provide an overview of the whole framework and introduce the details of each module in the subsequent sections.

\subsection{Framework Overview}

The goal of the visual sound separation task is to separate the sound components from the mixed signal by leveraging the visual information. Following previous works~\cite{Zhao_2018_ECCV,gao2019co,zhao2019sound}, we also adopt the `mix-and-separation' paradigm to carry out the training in a self-supervised manner.

Suppose we have two video clips $\{P, Q\}$ with corresponding audio signals~\{$x_P$, $x_Q$\}, the audio components are mixed to generate a synthetic mixture signal $x_m = x_P + x_Q$. For easy training, the mixed raw signal $x_m$ is first converted to the spectrogram $\mathbf{S_m}$ via Short Time Fourier Transform (STFT). The vision analysis network extracts the visual feature $\mathbf{f_{i}^{v}}$ ($i \in \{P, Q\}$) from the input frames for each video clip, while the audio feature $\mathbf{f_a}$ is generated by feeding the mixed spectrogram $\mathbf{S_m}$ into the audio analysis network. Afterward, the audio feature is fused with the visual features $\{\mathbf{f_{P}^{v}}, \mathbf{f_{Q}^v}\}$, respectively, to produce the separation masks $\{\mathbf{M_P}, \mathbf{M_Q}\}$. Finally, we multiply the mixed spectrogram with the predicted masks to obtain the clean spectrograms and produce the clean signals via Inverse STFT. 

Different from the original setting, 
we aim to explore a more challenging scenario to separate the unknown musical instruments. To achieve the adaptation ability on the novel categories, we introduce an additional consistency analysis network, which requires the predicted separation results to maintain both the \textbf{inter-modal} and \textbf{intra-modal} consistency. The inter-modal consistency is implemented with the synchronization of video and the corresponding separated audio~\cite{korbar2018cooperative,owens2018audio}, where the network can capture the audio-visual correlations when encountering new categories and acquire stronger transferring ability. Besides, based on the observation that instruments of the same type normally have similar timbres and tones, we add the intra-modal consistency supervision to the system, which will shorten the distance of the audio features from the same category and enlarge that from different categories in the embedding space.
Inspired by the fact that humans may spend more time observing and exploring repeatedly when encountering unfamiliar objects, we introduce the online matching mechanism during the inference stage so that the model can better fit the new instrument category. Specifically, the framework will make explicit adjustments for each sample pair by recurrently updating the model parameters from the supervision of the consistency loss. Please note that~\textit{no} Ground-Truth information is required since we only adopt the consistency loss as the error signal. The initial separation signals may be coarse due to the considerable gap between the training domain and the testing domain. But as the model becomes more familiar with the test sample, it can grasp more precise information and hence generate more delicate audio components. The experimental results demonstrate the effectiveness of the online matching strategy.

\subsection{Separation Network}

\begin{figure}
    \subfigure[]{
      \centering
      \label{fig_sep_network}
      \includegraphics[width=0.5\linewidth]{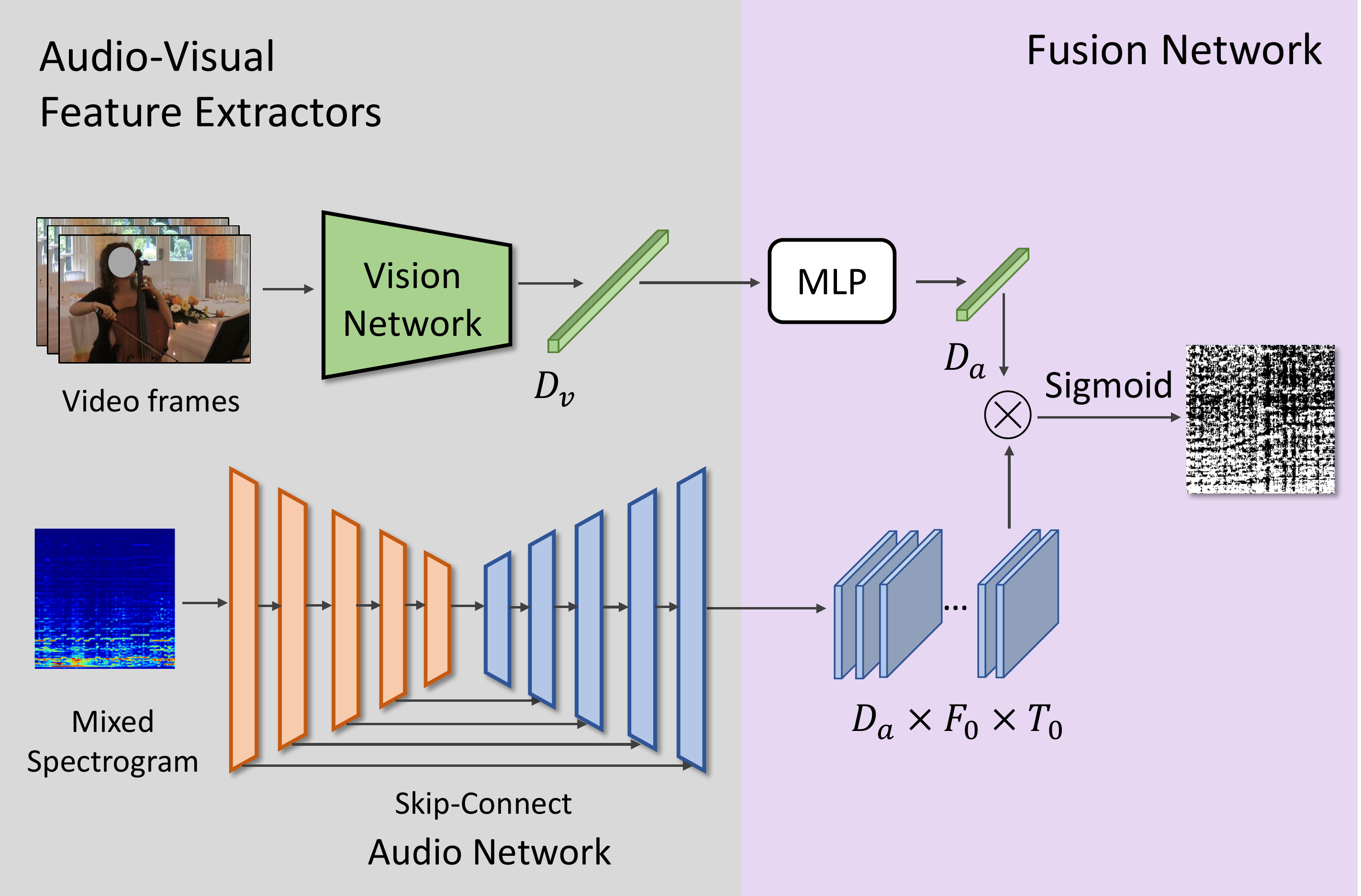}}
    \subfigure[]{
      \centering
      \label{fig_consistency}
      \includegraphics[width=0.46\linewidth]{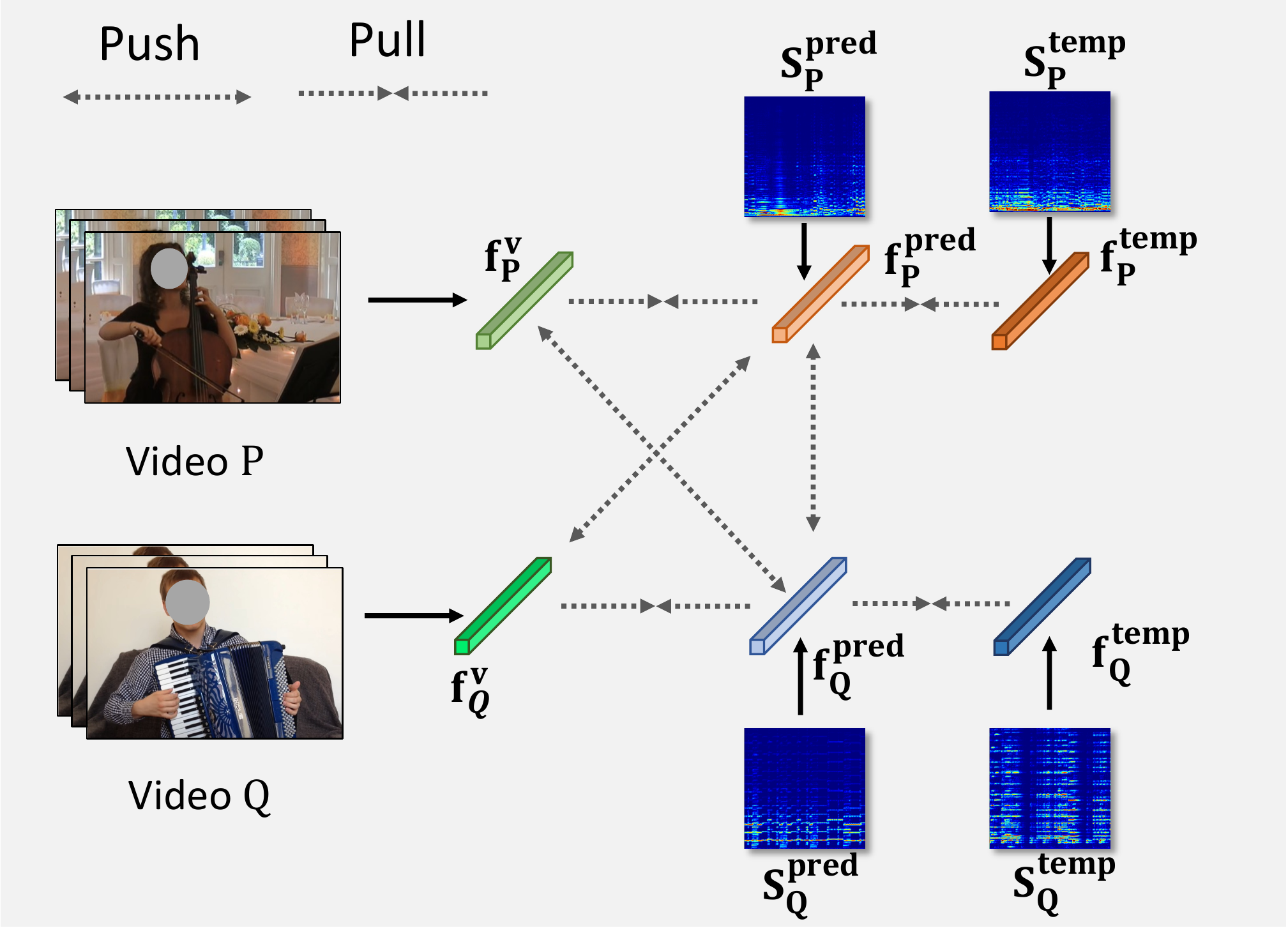}}
    \caption{(a) Structure of the separation network. 
    (b) Constraint relationships between the embedding pairs. For simplicity, we do not distinguish the inter-modal and intra-modal constraints but show the overall consistency relationships}
    \label{fig:my_label}
\end{figure}

The separation network is composed of three components, that is, the vision analysis network, the audio analysis network, and a fusion network. The mixed spectrogram passes through the audio analysis network to generate the audio feature. The visual feature is extracted by the vision analysis network for each video clip and then fused with the mixed audio feature in the fusion network to produce the separation mask. The process is illustrated in Fig.~\ref{fig_sep_network}. 

\subsubsection{Vision analysis network.} Videos contain rich visual cues, such as appearances, texture, motion, and so on. In our framework, we focus on motion information as the visual message for the following considerations. First, compared with the spatial semantics, motion relies less on category information, which makes it more effective guidance when dealing with novel classes. Second, humans can naturally associate the instrument playing actions with the sounds, regardless of the specific instrument category. Thus, exploiting such correlation will enhance the transferring ability of the system.

We adopt the fast pathway in the SlowFast network~\cite{feichtenhofer2019slowfast} as our vision analysis network to extract visual features. As the motion information is selected as the primary visual guidance, we remove the slow pathway and the lateral connection structure from the original SlowFast architecture and only keep the fast branch. Our vision network also preserves the high temporal resolution and low channel capacity properties, which can capture the detailed motions without introducing heavy parameter burdens. The specific implementation of the network will be provided in the supplementary. The vision network does not need optical flows and can directly learn the motion representations from the raw frames in an end-to-end manner.

\subsubsection{Audio analysis network.} Following previous works~\cite{Zhao_2018_ECCV,gao2019co}, we adopt a U-Net~\cite{ronneberger2015u} style encoder-decoder with skip-connections to extract the audio features. The U-Net consists of 5 downsampling convolution layers and 5 de-convolution layers for upsampling. The audio analysis network takes the mixed spectrogram $\mathbf{S_m}$ as input and yields audio feature of shape $D_a\times T_0 \times F_0$, where $T_0$ and $F_0$ refer to the temporal and frequency dimensions, respectively, and have the same values as $\mathbf{S_m}$. If not specified, we set $D_a = 64$ in the experiments.

\subsubsection{Fusion network} After the visual and audio features are extracted, we can fuse the visual guidance into the audio feature to compute the separation mask. Before the fusion, we adjust the channel dimension of the visual feature to $D_a$ via a linear projection and then apply a sigmoid activation on the projected feature. The activated visual feature is multiplied with the audio feature along the channel dimension to compute the fusion mask of shape $1\times T_0 \times F_0$. Finally, we activate the fusion mask via the sigmoid function to acquire the predicted separation mask. The separation loss $L_{mask}$ is the per-pixel binary cross-entropy loss between the predicted mask and the Ground-Truth mask. The Ground-Truth mask of each component is produced by checking whether the target spectrogram is dominant in the mixed spectrogram at every $T$-$F$ unit:

\begin{equation}
    M_i^{GT}(x,y) = [{S_i}(x,y) \geq {S_m}(x,y)], i\in\{P, Q\},
\label{mask_gt}
\end{equation}

\noindent where $(x,y)$ refers to the coordinates along the $T$-$F$ dimensions. 

\subsection{Consistency Network}
\label{consistency_loss_sec}

To raise the adaptation ability on novel classes, we propose two types of consistency contraints, i.e., inter-modal consistency and intra-modal consistency, which are both exerted on the predicted separation results. Thus, the mixed spectrogram $\mathbf{S_m}$ is multiplied by the predicted masks \{$\mathbf{M_P}$, $\mathbf{M_Q}$\} to develop the separated spectrograms \{$\mathbf{S^{pred}_P}$, $\mathbf{S^{pred}_Q}$\}. Furthermore, since the comparison of the raw spectrograms may not be very informative, we use high-level features to replace the low-level spectrograms for consistency computation, which are extracted from the consistency network. The network is stacked by 10 residual blocks, followed by a global max-pooling layer. The consistency embeddings of $\{\mathbf{S^{pred}_{P}}, \mathbf{S^{pred}_{Q}}\}$ are denoted as $\{\mathbf{f_{P}^{pred}}, \mathbf{f_{Q}^{pred}}\}$, respectively, which are both of 256-$d$ shape. The constraint relationships between the embedding pairs are depicted in Fig.~\ref{fig_consistency}. 

\subsubsection{Inter-modal consistency.}
Since the audio-visual associations in videos are natural and will not be disturbed by the category information, we add the inter-modal consistency to the system to strengthen the synchronization between the audio and visual elements so that it will present stronger adaptation ability with novel categories. Similar to~\cite{korbar2018cooperative}, the training objective is minimizing the distance on the positive pairs while enlarging the distance on the negative pairs. The positive pairs are synchronized audio-visual samples, i.e., the separated audio embeddings and their corresponding visual features $\{\mathbf{f_{i}^{pred}}, \mathbf{f_{i}^{v}}\},~i\in\{P, Q\}$. The negative pairs are obtained by cross-pairing the uncorrelated audio and visual features, that is, $\{\mathbf{f_{i}^{pred}}, \mathbf{f_{j}^{v}}\}$, $i\neq j, i,j\in \{P, Q\}$. 

At the beginning of training, the separation results may be poor since the network has not fully converged yet, and the suboptimal separation predictions may confuse the identification of positive pairs. Based on this consideration, we introduce the Ground-Truth audio features to assist the synchronization learning, which are extracted from $\{\mathbf{S_P}, \mathbf{S_Q}\}$ and denoted as $\{\mathbf{f_{P}^{GT}}, \mathbf{f_{Q}^{GT}}\}$. The loss weights between the predicted part and the Ground-Truth part vary according to the training time. The inter-modal consistency loss is defined as follows:

\begin{equation}
\begin{aligned}
    L_{inter-modal} & = \gamma(t)(D(\mathbf{f_{P}^{GT}}, \mathbf{f_{P}^v}) + D(\mathbf{f_{Q}^{GT}}, \mathbf{f_{Q}^v})) \\
    & + D(\mathbf{f_{P}^{pred}}, \mathbf{f_{P}^v}) + D(\mathbf{f_{Q}^{pred}}, \mathbf{f_{Q}^v}) \\
    & - D(\mathbf{f_{P}^{pred}}, \mathbf{f_{Q}^v}) - D(\mathbf{f_{Q}^{pred}}, \mathbf{f_{P}^v}),
\end{aligned}
\label{inter_modal_loss}
\end{equation}

\noindent where $D$ refers to the $L_2$ distance between two features and $\gamma(t)$ is the weight for the Ground-Truth assisted part that decays over training time. All features are normalized before computation.

\subsubsection{Intra-modal consistency.}
The design of the intra-modal consistency is based on two assumptions: (1) Instruments of the same category should have similar tones and timbres so their audio signals are supposed to be closer when projected to the feature space. (2) To achieve a higher quality separation result, the audio features of the two mixed videos should be pulled away. Please note that assumption (2) does not conflict with (1) because we require that the two mixed audios come from different instrument classes.

For the in-class similarity learning in assumption (1), we utilize audio signals $\{x_{P}^{temp}, x_{Q}^{temp}\}$ from the additionally sampled template video clips. Please note that the template clip comes from a different video of the same category as the separation target. The template audio signals are also converted to spectrograms via STFT and then pass through the consistency network to produce the high-level embeddings $\{\mathbf{f_{P}^{temp}}, \mathbf{f_{Q}^{temp}}\}$. The intra-modal consistency loss is shown as followed:

\begin{equation}
\begin{aligned}
L_{intra-modal} & = D(\mathbf{f_{P}^{pred}}, \mathbf{f_{P}^{temp}}) + D(\mathbf{f_{Q}^{pred}}, \mathbf{f_{Q}^{temp}}) \\
& - D(\mathbf{f_{P}^{pred}},\mathbf{f_{Q}^{pred}}),
\end{aligned}
\label{intra_modal_loss}
\end{equation}

\noindent where $D$ represents the $L_2$ distance between the features and all features are normalized before computation.
The consistency loss $L_{cs}$ is the sum of the inter-modal and intra-modal components:
\begin{equation}
\label{eq_consistency}
    L_{cs} = L_{inter-modal} + L_{intra-modal}
\end{equation}

Therefore, the overall loss function of our framework is:
\begin{equation}
    L = L_{mask} + \lambda L_{cs},
\label{total_loss}
\end{equation}
\noindent where $\lambda$ is the weight of consistency loss.

\subsection{Online Matching Strategy}

We introduce an online matching strategy to promote model compatibility with the samples from the novel domain in the inference phase. The parameters of networks will be fine-tuned explicitly for each sample pair by the backpropagation of error signals from the consistency loss. In this way, the online matching process can be regarded as `training during inference' but we only adopt the consistency loss as the supervision signal, and the optimization is based on one single pair. We emphasize that~\emph{no} Ground-Truth information is involved in this process so that the process can be regarded as a self-correction mechanism (the Ground-Truth assisted part in $L_{inter-modal}$ is excluded). 

For each sample pair, we optimize the model parameters via the consistency loss for several iterations and generate the refined separation masks for the pair based on the updated parameters. Before moving to the next pair, the parameters are switched to the original state so that the samples will not mutually affect each other. In practice, we fix all BatchNorm layers to avoid the fluctuations caused by the single sample input. Please refer to the supplementary for more details about the process. Our online matching strategy will not introduce any extra parameters but can bring consistent improvements.

\section{Experiments}
\subsection{Implementation Details}

Our pipeline is implemented with the PyTorch framework~\cite{paszke2019pytorch}. We use an Adam optimizer with betas (0.9, 0.999) and batch size 40. The weight of consistency loss $\lambda$ in Eq.~\ref{total_loss} is set to 0.01. The decay parameter $\gamma(t)$ in Eq.~\ref{inter_modal_loss} follows the function: $\gamma(t)$ = max(0.1, $0.9^{iter/100}$), where \textit{iter} refers to the training iterations. The framework is trained for 17000 iterations. The learning rate of the vision and consistency network are 1e-4 while the that of the audio and fusion network are 1e-3. During the online matching process, the learning rate is 1e-4 for the entire system and each sample pair is refined for 5 iterations. Since the consistency loss is the only supervision signal, we set $\lambda$ to 1.0 in this process. The data processing details will be provided in the supplementary. 

\subsection{Dataset and Evaluation Metrics}

We quantitatively evaluate our framework on the MUSIC-21 dataset~\cite{zhao2019sound} which contains 21 classes of instruments. The dataset is composed of untrimmed videos crawled from the YouTube website so that the contents are relatively diverse and complex. We randomly select 16 instruments as the training split: accordion, acoustic guitar, bagpipe, bassoon, cello, clarinet, congas, drum, erhu, flute, piano, pipa, tuba, ukulele, violin, and xylophone, while the other 5 classes as the testing split: banjo, electric bass, guzheng, saxophone, and trumpet. We denote this division as split-1 and conduct most experiments on it. But our method is not constrained to this split but proves to be effective on many other splits (See Sec.~\ref{ablation} for details). During training, two video clips from different categories in the training split are randomly selected and mixed. At the inference stage, we randomly pick up 300 pairs of video clips from the testing split to construct the testing dataset. In this way, the testing results are reliable and will not be disturbed by random combinations. 

We use the open-source mir\_eval library~\cite{raffel2014mir_eval} to conduct quantitative evaluations on the separated audios, where three metrics are selected:  Signal-to-Distortion Ratio (SDR), Signal-to-Interference Ratio
(SIR), and Signal-to-Artifact Ratio (SAR). The units are dB. The SDR score is normally regarded as the most convincing metric.

\subsection{Quantitative Results}
\label{quant_results}

The results of the baseline and our SeCo are shown in Table~\ref{main_results}. The traditional method NMF-MFCC~\cite{spiertz2009source} does not exhibit obvious degeneration on the testing splits, which is reasonable since it is non-learned. The traditional algorithm can only return unpaired separation signals so that we conduct the exclusive matchings and take the overall best results. Even with the best matching, it still presents trivial performances, which means that the traditional method lacks the potential for further improvements. The deep-learning based Sound-of-Pixels~\cite{Zhao_2018_ECCV} and Co-separation~\cite{gao2019co} methods only adopt the spatial semantics as visual cues and do not capture the temporal correlations. The results indicate that they fail to successfully separate the sounds of the novel classes. The spatial semantics such as appearances and textures are closely related to the category so the learned visual representations cannot provide sufficient separation guidance when encountering the novel classes. 
On the other hand, if the basic components of the visual guidance are transferred from spatial to temporal, i.e., the motion information, the over-fitting symptom can be alleviated, where we can see that motion only SeCo provides a relatively good baseline. Compared with the spatial semantics, the temporal information is less category-specific, which can serve as more effective separation guidance with the novel instrument types. Despite the progress, simply replacing the visual modality still fails to bring satisfying performance improvements.  In general, the baseline results demonstrate that it is a challenge for the normal frameworks to handle the sound separation task on musical instruments that have never seen before.

Our SeCo framework outperforms all baseline methods by a large margin under this challenging scenario, which demonstrates the effectiveness of our method. Although changing the visual modality can bring a relatively good starting point, we argue that the main improvements come from the consistency loss. By accomplishing the music separation on novel categories, our method outcomes the limitation of prior works and proves the feasibility of deploying a more general setting. The results may expand the scope of visual music separation and make the task more versatile.

\begin{table}
\caption{Sound separation results on the MUSIC-21 testing dataset, higher is better for all metrics. 
The SeCo (motion only) does not adopt the consistency loss and utilizes the motion information as visual guidance. SeCo incorporates both the consistency loss and the online matching strategy, which outperforms all baselines by a large margin}
\label{main_results}
\centering
\setlength{\tabcolsep}{4mm}{
\begin{tabular}{l|ccc}
\hline
Method & SDR & SIR & SAR \\
\hline
NMF-MFCC~\cite{spiertz2009source} & 0.90 & 5.37 & 6.94 \\
Sound-of-Pixels~\cite{Zhao_2018_ECCV} & -2.56 & 2.42 & 4.97 \\
Co-Separation~\cite{gao2019co} & -2.89 & 1.97 & 5.23 \\
SeCo (motion only) & 1.16 & 4.39 & 9.64 \\
SeCo & 4.01 & 7.13 & 11.62 \\
\hline
\end{tabular}}
\end{table}

\subsection{Ablation Study}
\label{ablation}
\subsubsection{Inter-modal v.s. intra-modal consistency.} We conduct experiments to investigate the importance of the different loss components and report the results in Table~\ref{loss_ablation}. We can see that both the inter-modal consistency and the intra-modal consistency will promote the separation performance on novel musical instrument types, since adopting either loss will win the baseline method (\textit{w.o.} the consistency loss). As depicted in the Table, we achieve the best results by employing both inter-modal and intra-modal consistency losses regarding all evaluation criteria.

\begin{table}
\caption{Ablation study of the importance of consistency loss components on the MUSIC-21 testing dataset. The results show that both inter-modal and intra-modal consistency loss contribute to the separation performance on novel categories}
\label{loss_ablation}
\centering
\begin{tabular}{cc|ccc}
\hline
$L_{inter-modal}$ & $L_{intra-modal}$ & SDR & SIR & SAR \\
\hline
\scriptsize{\XSolidBrush} & \scriptsize{\XSolidBrush} & 1.16 & 4.39 & 9.64 \\
\scriptsize{\CheckmarkBold} & \scriptsize{\XSolidBrush} & 2.05 & 4.93 & 10.48 \\
\scriptsize{\XSolidBrush} & \scriptsize{\CheckmarkBold} & 2.16 & 5.40 & 10.01 \\
\scriptsize{\CheckmarkBold} & \scriptsize{\CheckmarkBold} & \textbf{2.37} & \textbf{5.03} & \textbf{11.29} \\
\hline
\end{tabular}
\end{table}

\subsubsection{Comparison of different visual cues.} To investigate the effects of visual cues, we conduct experiments on three visual modalities, i.e., image, skeleton, and motion. Implementation details of the image and skeleton are provided in the supplementary. 

The performance comparisons of different visual modalities are summarized in Table~\ref{modality_table}, where both the baseline and our SeCo approaches are presented. From the baseline results, we can see that when using the static images as the visual guidance, the model fails to successfully separate sounds from the unknown musical instruments. We suspect that the failure may come from the dependence of spatial information on categories, which causes over-fitting to the training scenarios, as analyzed in Sec.~\ref{quant_results}. The visual features of the skeleton modality incorporate both the spatial and temporal relations and we can see that the over-fitting problem has been a little bit alleviated. However, simply replacing the images with skeletons is not enough to generate the optimal results. The possible reason is that the skeleton data only retain the joint coordinates of the players while discarding much detailed information in the original video clips. Such simple and intuitive visual cues may hinder the ability to conduct visual sound separation on new categories, given no additional prior knowledge. In contrast, for the motion modality, the 3d-CNN based vision analysis network directly learns the temporal representations from the original video clips, which can capture richer semantics. This property makes the motion modality better visual cues in our setting and provides an advanced starting point for further improvements.

In addition to the analysis of the baseline results, we also evaluate the performances of our SeCo framework when utilizing different vision modalities. Please note that the inter-modality loss is not applicable to the image-based visual cues. Therefore, for a fair comparison, only the intra-modality loss is applied as the consistency constraints for all modalities. The SeCo pipeline includes both the normal training and the online matching process and we can find that SeCo considerably exceeds the baseline on all three modalities. The results verify the robustness and flexibility of our approach. 

\begin{table}
\centering
\caption{Sound separation results when utilizing visual cues of different modalities. We report results from both the baseline and the SeCo method. The SeCo method includes the normal training and the subsequent online matching process}
\label{modality_table}
\small
\begin{tabular}{l|ccc|ccc}
\hline
\multirow{2}{*}{} &
\multicolumn{3}{c|}{Baseline} & \multicolumn{3}{c}{SeCo (\textit{w.} O.M.)} \\
\cline{2-7}
  & SDR & SIR & SAR & SDR & SIR & SAR \\
\hline
Image & -2.56 & 2.42 & 4.97 & 2.95 & 6.34 & 9.45 \\
Keypoints & -1.35 & 2.83 & 6.05 & 3.43 & 7.82 & 9.97 \\
Motion & \textbf{1.16} & \textbf{4.39} & \textbf{11.10} & \textbf{3.91} & \textbf{6.50} & \textbf{11.34}  \\
\hline
\end{tabular}
\end{table}

\subsubsection{Division of musical instrument categories.} 
To ensure that our SeCo framework does not rely on certain instrument types, we make verifications on different train/test splits.
These extra train/test splits also follow the 16/5 category division but the internal instrument types vary from each other. We conduct experiments on 2 additional splits and list the results in Table~\ref{splits_table}, which demonstrate that our SeCo framework is effective on various splits rather than constrained to certain specific instrument types. The category divisions of the splits are provided in the supplementary. Moreover, the results also confirm the robustness of our online matching strategy, which can also handle different instrument types.

\begin{table}[ht]
\begin{center}
\caption{Separation results on different train/test splits. 
We show results~\textit{w/o} and~\textit{w.} the online matching strategy (denoted as O.M. in table), respectively}
\label{splits_table}
\begin{tabular}{c|ccc|ccc}
\hline
& \multicolumn{3}{c|}{\textit{w/o} O.M.} & \multicolumn{3}{c}{\textit{w.} O.M.} \\
\cline{2-7}
 & SDR & SIR & SAR & SDR & SIR & SAR \\
\hline
Split-1 & 
2.37 & 5.03 & 11.29 & 4.01 & 7.13 & 11.62 \\

Split-2 & 
3.81 & 6.28 & 12.67 & 5.72 & 8.87 & 13.38 \\
Split-3 & 
2.72 & 5.23 & 12.04 & 3.89 & 6.93 & 12.50 \\
\hline
\end{tabular}
\end{center}
\end{table}

\subsubsection{Online matching iterations.} 
\begin{figure}[h]
    \centering
    \includegraphics[width=0.65\textwidth]{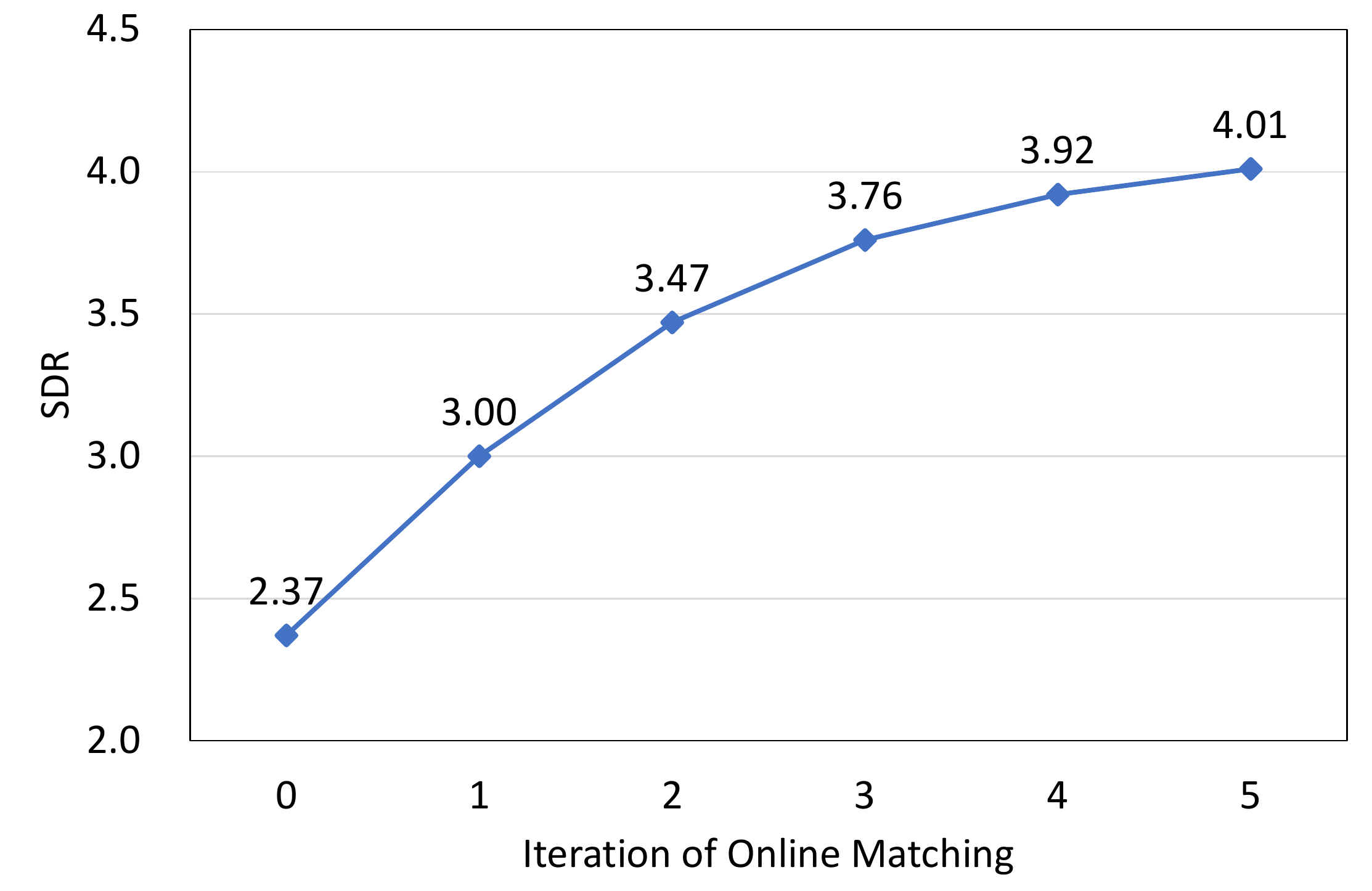}
    \caption{Trend of the SDR scores with different online matching iterations. Iteration 0 refers to the result of not adopting the online matching strategy}
    \label{om_iterations}
\end{figure}

The key hyper-parameter of the online matching strategy is the number of optimization steps at each sample pair, denoted as~\textit{iterations}. We investigate the influence of changing the optimization iterations and visualize the trends in Fig.~\ref{om_iterations}. We use the SDR scores to represent the performances since it is the most important evaluation metric. Naturally, increasing the iterations will help the network get more familiar with the current sample and thus produce better separation results but we can also observe the marginal effect with longer iterations. The trend also indicates that the online matching strategy can bring stable improvements instead of random fluctuations. Please note that the online matching strategy will only update the existing parameters so that the performance gains are obtained at no cost of extra parameters.

\subsection{Qualitative Results}
\begin{figure}[ht]
\centering
\includegraphics[width=0.8\textwidth]{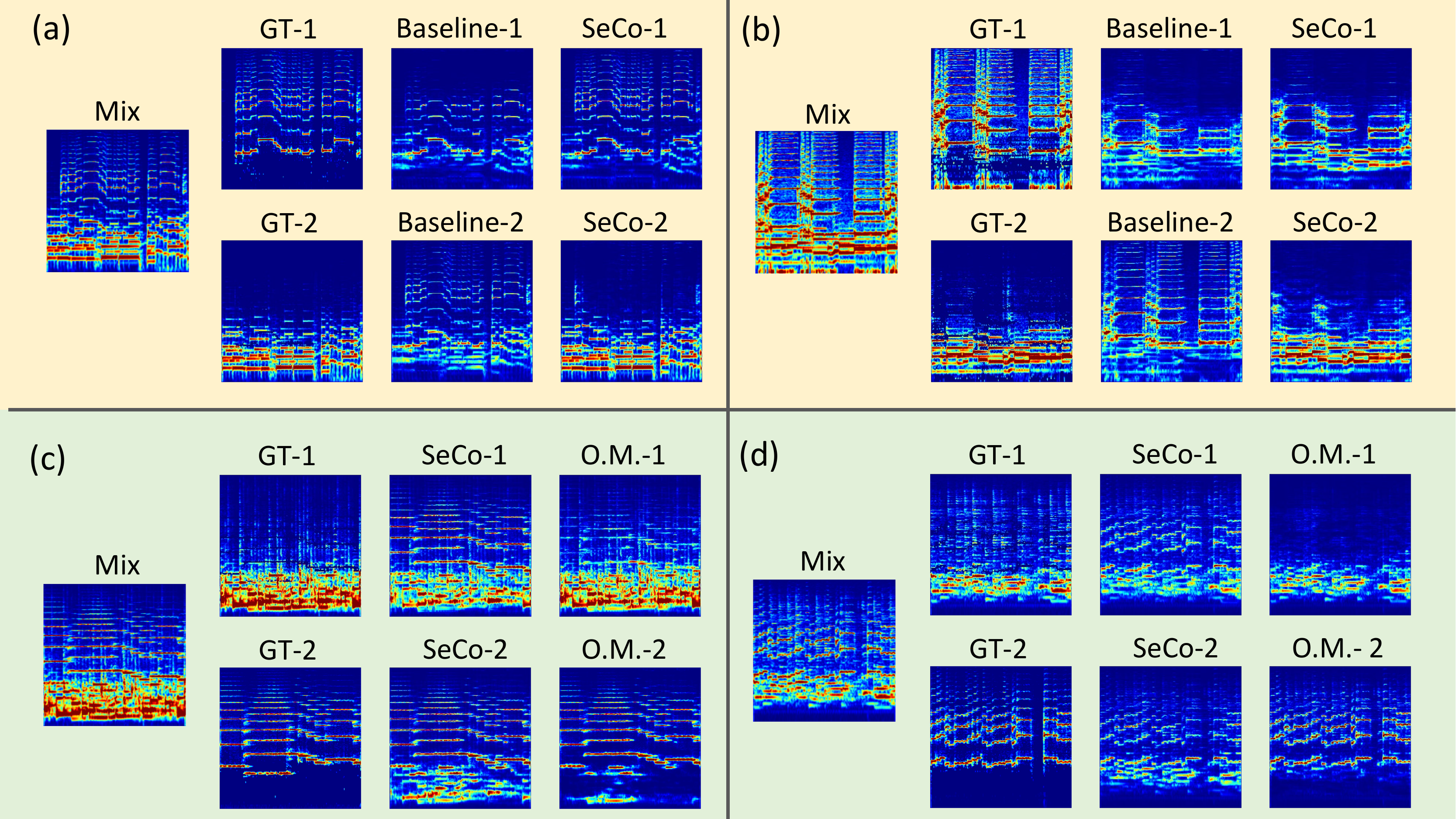}
\caption{Visualization of the separated spectrograms on the MUSIC-21 testing dataset. The index `1 \& 2' refers to the two audio components to be separated and GT stands for the 'Ground-Truth' spectrograms}. 
\label{fig_vis}
\end{figure}

We visualize four cases of the separated spectrograms on the MUSIC-21 testing dataset in Fig.~\ref{fig_vis}. In (a) and (b), we compare the baseline method and the SeCo method. Both methods adopt the motion information as visual cues but the consistency loss is not included in the baseline method. We can observe that the baseline results lose many details and contain components from its mixture audio counterpart, while the SeCo results are closer to the Ground-Truth spectrograms. The comparisons vividly show the effectiveness of the consistency loss. 

Although our SeCo method is superior to the baseline method, it may still encounter the detail missing and noisy problems due to the challenge of the unknown musical sound separation task. However, these problems can be alleviated by the subsequent online matching process. As shown in (c) and (d), the online matching process can correct the undesirable effects from the other audio component and grasp more details. In this way, we can obtain separation results of higher quality. 

\section{Conclusions}
In this work, we explore a novel and challenging scenario of visual sound separation, i.e., music separation on unknown musical instruments. To promote the adaptation ability for the deep model on unfamiliar melodies, we design the Separation-with-Consistency~(SeCo) framework that utilizes both the inter-modal and intra-modal consistency constraints. Moreover, to fully exploit the consistency potentials, we devise the online matching strategy, which further boosts the system performance with no extra parameter costs. We conduct extensive ablation studies to analyze the key factors in the system, which also exhibit that our SeCo framework is effective and robust on various visual modalities and musical instrument types. Our work proves the feasibility of separation on novel musical instruments and hence expands the scope of the visual sound separation task. We wish our work could inspire the community to further explore the transferability of deep models in the audio-visual learning field.

%
%
\bibliographystyle{splncs04}
\bibliography{egbib}
\end{document}


\pagestyle{headings}
\mainmatter
\def\ECCVSubNumber{3030}  

\title{Supplementary Material} 

\titlerunning{ECCV-22 submission ID \ECCVSubNumber} 
\authorrunning{ECCV-22 submission ID \ECCVSubNumber} 
\author{Anonymous ECCV submission}
\institute{Paper ID \ECCVSubNumber}

\maketitle

\appendix
\renewcommand\thefigure{A\arabic{figure}}
\renewcommand\thetable{A\arabic{table}}
\renewcommand\thealgorithm{A\arabic{algorithm}}
\setcounter{figure}{0}
\setcounter{table}{0}
\setcounter{algorithm}{0}

\section{Architecture of the Vision Analysis Network}

In Table.~\ref{tab_vision_net}, we show the architecture details of the vision analysis network. The kernel shape is denoted as $\{T\times{S_z}^2, C\}$ for temporal, spatial and channel dimensions, respectively. We do not perform the downsampling operation on the temporal dimension in the entire structure so that the temporal stride is fixed to 1.  The parameter $\alpha$ in the bottleneck structure represents the channel reduction ratio and is set to $1/4$.
\begin{table}[]
\caption{Implementation details of the vision analysis network. `Stack' refers to the number of stacked bottleneck blocks in each residual stage and `Stride' to the stride value along the spatial dimensions for the conv3d or pooling operations}
\label{tab_vision_net}
\centering
\begin{tabular}{c|cccc}
\hline
Structure & \multicolumn{4}{c}{Information} \\
\hline
\multirow{3}*{Bottleneck} &
\multicolumn{4}{c}{\{3\times$1^2$, $\alpha C$\} Conv3d, BN, ReLU} \\
~ & \multicolumn{4}{c}{\{1\times$3^2$, $\alpha C$\} Conv3d, BN, ReLU} \\
~ & \multicolumn{4}{c}{\{1\times$1^2$, $C$\} Conv3d, BN} \\
\hline
\multirow{9}*{Vision Network} & Stage & Operator & Stack & Stride \\
\cline{2-5}
~ & Input & - & -  & - \\
~ & Conv1 & \{5\times $7^2$, 8\} & - & 2 \\
~ & Pool1 & 1\times$3^2$, MaxPool3d & - & 2 \\
~ & Res1 & Bottleneck, $C$=32 & 3 & 1 \\
~ & Res2 & Bottleneck, $C$=64 & 4 & 2 \\
~ & Res3 & Bottleneck, $C$=128 & 6 & 2 \\
~ & Res4 & Bottleneck, $C$=256 & 3 & 2 \\
~ & Pool2 & Global AvgPool3d & - & - \\
\hline
\end{tabular}
\end{table}

\section{Algorithm for Online Matching Strategy}

In Alg.~\ref{alg_om}, we illustrate the optimization process of the online matching strategy. 

\begin{algorithm}
	\begin{algorithmic}[1]
	\caption{}
	\label{alg_om}
	\Require $\mathcal{P}$: the testing dataset
	\Require $\theta_0$: Pre-trained model parameters
	\Require $\mathcal{T}$, $\beta$: hyper-parameters for online matching
	\For{$p_k$ in $\mathcal{P}$}
	\Comment{$p_k$ refers to single test pair}
	\For{$\tau$ = $1$ to $\mathcal{T}$}
	\State Compute the gradient $\bigtriangledown_{\theta}L_{cs}$ using $\theta_{\tau-1}$ and $p_k$
	\State Update the parameters: $\theta_{\tau} = \theta_{\tau-1} - \beta\bigtriangledown_{\theta}L_{cs}$
	\EndFor
	\State Compute the refined separation masks:
	\Statex \qquad$M_k=Model(p_k|\theta_{\mathcal{T}})$
	\State Reset model parameters to $\theta_{0}$
	\EndFor
	\end{algorithmic}
	\Return Refined separation masks for all pairs in $\mathcal{P}$
\end{algorithm}

\section{Data Processing Details of SeCo}

During training, we randomly segment a 6-second video clip from the dataset. The audio signal is re-sampled to 11kHz. The audio signals are transformed to spectrogram by STFT with window size 512 and hop length 256. Thus, we obtain a Time-Frequency audio representation of shape 512$\times$256. The spectrograms are further re-sampled on a log-frequency scale to produce 256$\times$256 T-F representations as the U-Net input.

For the visual side, we take 24 frames for every video clip, whose resolution is 128$\times$128. During training, the frames are resized to 160$\times$160 and a region of 128$\times$128 is randomly cropped as input. Other augmentation operations include horizontally flipping at 0.5 probability and normalization. At the inference stage, the randomly cropping operation is replaced with central cropping and the horizontal flip is skipped to reduce the effects of random factors.

\section{Implementation Details on the Image and Skeleton Modality}

For the image modality, we use a ResNet-18~\cite{he2016deep} network to extract the visual features after the $4^{th}$ ResNet stage. The convolution operations are replaced with the dilated convolution with the dilation set to 1, as in~\cite{Zhao_2018_ECCV}. Finally, a global MaxPooling operation is applied to generate the visual feature vector. In this way, the features are only convolved on the spatial dimensions while the temporal representations are not learned. 

For the skeleton modality, we use the pretrained OpenPose toolbox~\cite{cao2017realtime} instead of the manual annotations to estimate the 2D coordinates and the confidence scores for keypoints of the players in the video clips. Specifically, we obtain 18 joints for the human body and 21 joints for each hand. We select the Spatial-Temporal Graph CNN (ST-GCN)~\cite{yan2018spatial} as the vision analysis network to effectively extract features from the skeleton inputs. We follow~\cite{cao2017realtime} to build the edges between keypoints and form the spatial graph. The ST-GCN is composed of 11 layers with residual connections. After passing through the ST-GCN backbone, a global average-pooling operation is exerted on the feature map to produce a visual feature vector. For fair comparison of different modalities, we only adopt the skeleton modality as the visual cues instead of combining the skeleton and appearance information, which is different from~\cite{gan2020music}.

\section{Category Distribution of Splits}

In Table.~\ref{table_class}, we list the categories in the testing set for different splits. The other 16 categories will serve as the training set.

\begin{table}[]
    \centering
     \caption{Category distribution of different splits. We list the 5 categories for testing while the remaining 16 categories serve the training set}
    \label{table_class}
    \begin{tabular}{l|c}
    \hline
    Split &  Test Class\\
    \hline
    Split-1 & banjo, electric bass, guzheng, saxophone, trumpet \\
    Split-2 & bagpipe, drum, flute, piano, saxophone \\
    Split-3 & congas, guzheng, saxophone, trumpet, ukulele \\
    \hline
    \end{tabular}
\end{table}
%
%
\bibliographystyle{splncs04}
\bibliography{egbib}